# Distinctive momentum dependent charge-density-wave gap observed in CsV$_3$Sb$_5$ superconductor with topological Kagome lattice

Zhengguo Wang[1,†], Sheng Ma[2,3,†], Yuhang Zhang[2,3], Haitao Yang[2,3,4,6,*], Zhen Zhao[2,3], Yi Ou[1], Yu Zhu[1], Shunli Ni[2,3], Zouyouwei Lu[2,3], Hui Chen[2,3,4,6], Kun Jiang[2], Li Yu[2,3,6,*], Yan Zhang[1,*], Xiaoli Dong[2,3,6], Jiangping Hu[2,5], Hong-Jun Gao[2,3,4,6], and Zhongxian Zhao[2,3,6]

[1]International Centre for Quantum Materials, School of Physics, Peking University, Beijing 100871, China
[2]Beijing National Laboratory for Condensed Matter Physics and Institute of Physics, Chinese Academy of Sciences, Beijing 100190, China
[3]University of Chinese Academy of Sciences, Beijing 100049, China
[4]CAS Center for Excellence in Topological Quantum Computation, University of Chinese Academy of Sciences, Beijing 100190, PR China
[5]Kavli Institute of Theoretical Sciences, University of Chinese Academy of Sciences, Beijing, 100190, China
[6]Songshan Lake Materials Laboratory, Dongguan, Guangdong 523808, China

†These authors contributed equally to this work

*E-mail: li.yu@iphy.ac.cn (L. Y.); htyang@iphy.ac.cn (H. Y.); yzhang85@pku.edu.cn (Y. Z.);

**CsV$_3$Sb$_5$ is a newly discovered Kagome superconductor that attracts great interest due to its topological nontrivial band structure and the coexistence of superconductivity and charge-density-wave (CDW) with many exotic properties. Here, we report the detailed characterization of the CDW gap in high-quality CsV$_3$Sb$_5$ single crystals using high-resolution angle-resolved photoemission spectroscopy. We find that the CDW gap is strongly momentum dependent. While gapped around the *M* point, the electronic states remain gapless around the Γ point and along the Γ-*K* direction. Such momentum dependence indicates that the CDW is driven by the scattering of electrons between neighboring *M* points, where the band structure hosts multiple saddle points and the density of state diverges near the Fermi level. Our observations of the partially gapped Fermi surface and strongly momentum-dependent CDW gap not only provide a foundation for uncovering the mechanism of CDW in CsV$_3$Sb$_5$, but also shed light on the understanding of how the CDW coexists with superconductivity in this topological Kagome superconductor.**

Kagome lattice possesses naturally both Dirac band dispersion and flat band, providing an ideal platform for realizing novel electron states with distinctive properties [1,2]. Recently, a new family of Kagome material AV$_3$Sb$_5$ (A = K, Rb, Cs) has been discovered and draws great attentions [3-6]. Its crystal structure consists of an alternative stacking of the intercalated alkali-metal layer and the V$_3$Sb$_5$ layer that contains an ideal two-dimensional Kagome lattice of V atoms coordinated by Sb atoms [Fig. 1(a)]. The previous density functional theory (DFT) calculations and angle-resolved photoemission spectroscopy (ARPES) experiments show consistent results, suggesting that the AV$_3$Sb$_5$ is a Z$_2$ topological metal with topological surface states [3-6]. Following studies show that superconductivity (SC) emerges in AV$_3$Sb$_5$ with the critical temperature ($T_c$) up to 3.5 K [4,6-9]. More intriguingly, a giant anomalous Hall effect is observed [5,10], while surprisingly no long-range magnetic order has been observed from the magnetic susceptibility, neutron scattering, and muon spin resonance (μSR) measurements[3,5,6,11]. All these properties make AV$_3$Sb$_5$ an ideal system to study the interplay among the superconductivity, topological nontrivial band structure, and other rich phenomena within the Kagome lattice.

Besides all above, one key characteristic of AV$_3$Sb$_5$ is that it enters a charge-density-wave (CDW) state prior to the superconducting transition [3,4,6-8,10-17]. The resistivity and specific heat measurements show an anomaly at $T^*$ ranging from 80 K to 110 K [3,4,6,8]. The scanning tunneling microscopy (STM) and X-ray diffraction measurements confirm that below $T^*$ a CDW forms with a 2×2 star of David deformation of Kagome lattice [4,7,12,14,17]. STM measurement shows possible existence of chiral charge order at the low temperature [12]. Theoretical calculation also proposes that the CDW in AV$_3$Sb$_5$ is a chiral flux phase that breaks the time-reversal symmetry [15,16]. Such an exotic CDW phase is not only connected with the anomalous Hall effect [10,12,15,16], but also imply the existence of an unconventional superconductivity in AV$_3$Sb$_5$ with distinctive pairing symmetries [9,13,14,18,19] and topological character [20].

CDW is commonly believed to originate from the Peierls instability of electrons at the Fermi level ($E_F$). This instability is reflected in a gap opening on the Fermi surface in a CDW state. Therefore, for understanding the mechanism of a CDW state, delineating the CDW gap distribution in momentum space is essential. The previous STM work of AV$_3$Sb$_5$ observed a 20 meV energy gap at the low temperature and attributed it to a CDW gap [12,14,20]. However, previous APRES studies failed to resolve any CDW gap opening below $T^*$ presumably due to poor sample quality or surface inhomogeneity [4,5,7,21]. Therefore, it is still an open question that how the CDW gap develops on the Fermi surface of AV$_3$Sb$_5$ that consists of multiple Fermi pockets in different momentum locations. Here, we studied the CDW gap distribution of CsV$_3$Sb$_5$ using high-resolution ARPES. We resolve a clear CDW gap opening below $T^*$. Our detailed gap mapping shows that the CDW gap of CsV$_3$Sb$_5$ is strongly momentum dependent. In the CDW state, gap opens on the sections of Fermi surface around the Brillouin zone boundaries (M) while leaving the other sections of Fermi surface ungapped. Such gap distribution indicates that the scattering of electrons between neighboring M points plays a dominating role in driving the CDW in AV$_3$Sb$_5$. These findings provide crucial clues to the formation of CDW and a better understanding of its coexistence with superconductivity in AV$_3$Sb$_5$.

High quality single crystals of CsV$_3$Sb$_5$ were synthesized using the self-flux method [3], and the superconducting transition temperature was characterized as $T_c$ ~3.5 K [9]. The CDW transition temperature $T^*$ is determined to be 94 K from the clearly visible kink in the resistivity measurement [Fig. 1(b)]. ARPES measurements were performed on a home-made ARPES system at Peking University equipped with a DA30L electron analyser and a helium discharging lamp. The photon energy is 21.218 eV. The overall energy resolution was ~10 meV and the angular resolution was ~0.3°. The crystals were cleaved *in-situ* and measured in vacuum with a base pressure better than 6 × 10$^{-11}$ mbar. The Fermi level for the samples was referenced to that of a gold crystal attached onto the sample holder by Ag epoxy.

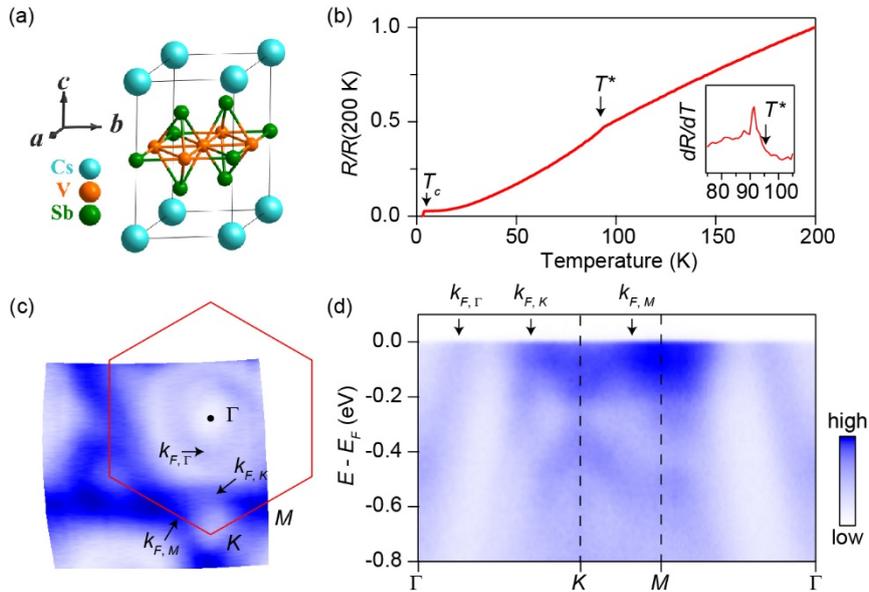

**Figure 1.** (a) Schematic drawing of the lattice structure of CsV$_3$Sb$_5$. (b) Temperature dependence of the normalized resistivity taken in CsV$_3$Sb$_5$. The inset is the first derivative of resistivity versus temperature. (c) Fermi surface mapping and (d) high symmetry band dispersion of CsV$_3$Sb$_5$ taken at 15 K. The Fermi crossings ($k_{F,\Gamma}$, $k_{F,K}$, and $k_{F,M}$) are pointed out by the black arrows.

The measured Fermi surface and band structure of $CsV_3Sb_5$ are shown in Figs. 1(c) and 1(d). The results are consistent with the previous ARPES studies and DFT calculation [3-5]. The Fermi surface consists of a circular-like electron pocket at the Brillouin center ($\Gamma$) and multiple triangle-like pockets around the Brillouin corner ($K$). In the energy-moment cuts taken along the high-symmetry directions [Fig.1 (d)], the Fermi crossings around the $\Gamma$ and $K$ points ($k_{F,\Gamma}$ and $k_{F,K}$) are clearly resolved. According to the band calculation [3], the band structure of $CsV_3Sb_5$ hosts multiple saddles points at the $M$ point. As a result, the observed band dispersion flattens near the $M$ point, forming an accumulation of density of state at the Fermi crossing near the $M$ point ($k_{F,M}$).

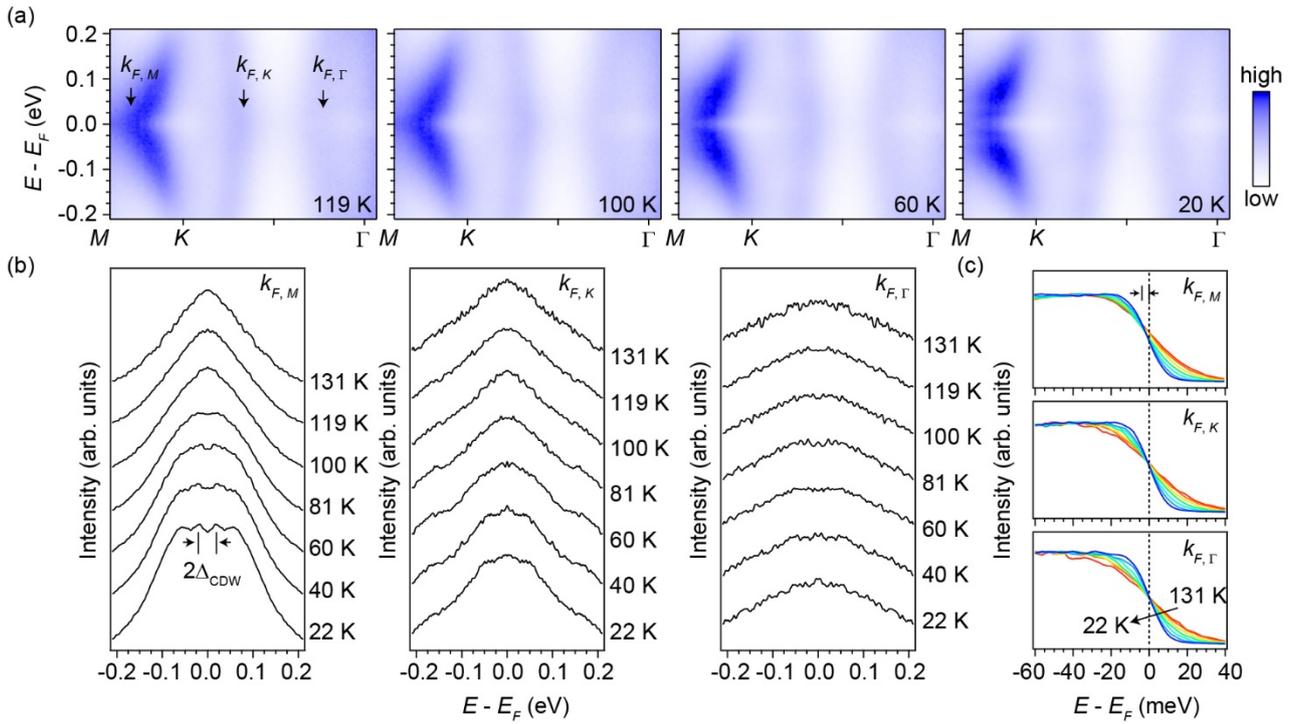

**Figure 2.** (a) Temperature dependence of the energy-momentum cut taken along the $\Gamma$-$K$-$M$ direction. The data were symmetrized with respect to $E_F$ using the standard symmetrization method [22]. The black arrows point out the Fermi crossings ($k_{F,\Gamma}$, $k_{F,K}$, and $k_{F,M}$). (b) Temperature dependence of the symmetrized energy-distribution curves (EDCs) taken at $k_{F,\Gamma}$, $k_{F,K}$, and $k_{F,M}$. (c) Temperature dependence of the raw EDCs taken at $k_{F,\Gamma}$, $k_{F,K}$, and $k_{F,M}$.

The temperature dependent experiment has also been performed along the $\Gamma$-$K$-$M$ direction, and the results are shown in Fig. 2. Note that the CDW gap is not necessarily particle-hole symmetric. We symmetrized the spectra with respect to $E_F$ only for better visualizing the energy gap opening. There is no energy gap at temperatures above $T^*$. At $k_{F,\Gamma}$, $k_{F,K}$, and $k_{F,M}$, the symmetrized spectra are characterized by a gapless behavior showing sharp peaks centering at $E_F$ [Figs. 2(a) and 2(b)]. Upon entering the CDW state, the symmetrized spectra at $k_{F,M}$ show a clear gap opening behavior. One peak splits into two peaks forming a dip-like feature at $E_F$. And then, another peak at higher energy can be better resolved at lower temperatures. The gap opening can also be visualized in the raw spectra as characterized by a shift of leading edge to higher binding energy [Fig. 2(c)]. If we assume that the energy gap is particle-hole symmetric, we could obtain a gap magnitude ($\Delta_{cdw}$) that is around 20 meV, which is well consistent with the STM measurement [12,20]. In contrast to the gap opening behavior observed at $k_{F, M}$, the symmetrized spectra at $k_{F, \Gamma}$ and $k_{F, K}$ show gapless behavior in the CDW state. Consistently, there is no leading-edge shift in the raw spectra taken at $k_{F, \Gamma}$ and $k_{F, K}$ [Fig. 2(c)].

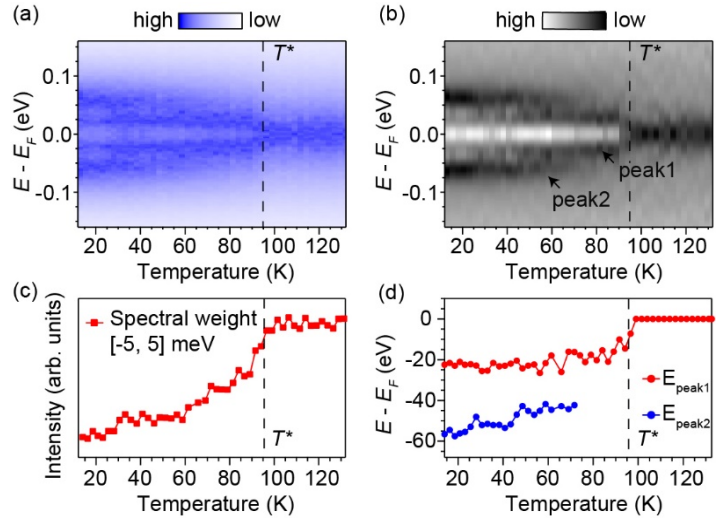

**Figure 3.** The merged image of the temperature dependence of (a) symmetrized EDCs and (b) the symmetrized EDC curvatures taken at $k_{F, M}$ near the $M$ point. Note that, in the curvature image, the minima represent the peak positions. (c) The temperature dependence of the spectral weight that is integrated with in a 10 meV energy window at $E_F$. (d) The temperature dependence of the energy positions of the low-energy peak (peak1) and the high-energy peak (peak2). The peak positions are determined by fitting the local maximum position with a Gaussian peak.

To further characterize how the energy gap opens in the CDW state, we measured the detailed temperature dependence of the spectrum taken at $k_{F, M}$ near the $M$ point (Fig. 3). The symmetrized energy distribution curves (EDCs) and their curvatures are merged into images to better visualize the evolution of spectra through the CDW transition [Fig. 3(a) and 3(b)]. To quantify the gap opening, we integrated the spectral weight within a 10 meV energy window at $E_F$ and determined the peak (peak1 and peak2) positions by fitting the local maximum position with a Gaussian function [Fig. 3(c) and 3(d)]. The temperature dependences of the spectral weight and the peak1 position all show an order-parameter-like behavior below $T^*$, indicating that the observed energy gap is indeed a CDW gap that characterizes the CDW order parameter. Note that, the peak2 is difficult to be distinguished above ~70 K, which may be due to the peak broadening, spectral weight losing, or the merging of peak1 and peak2 near the CDW transition.

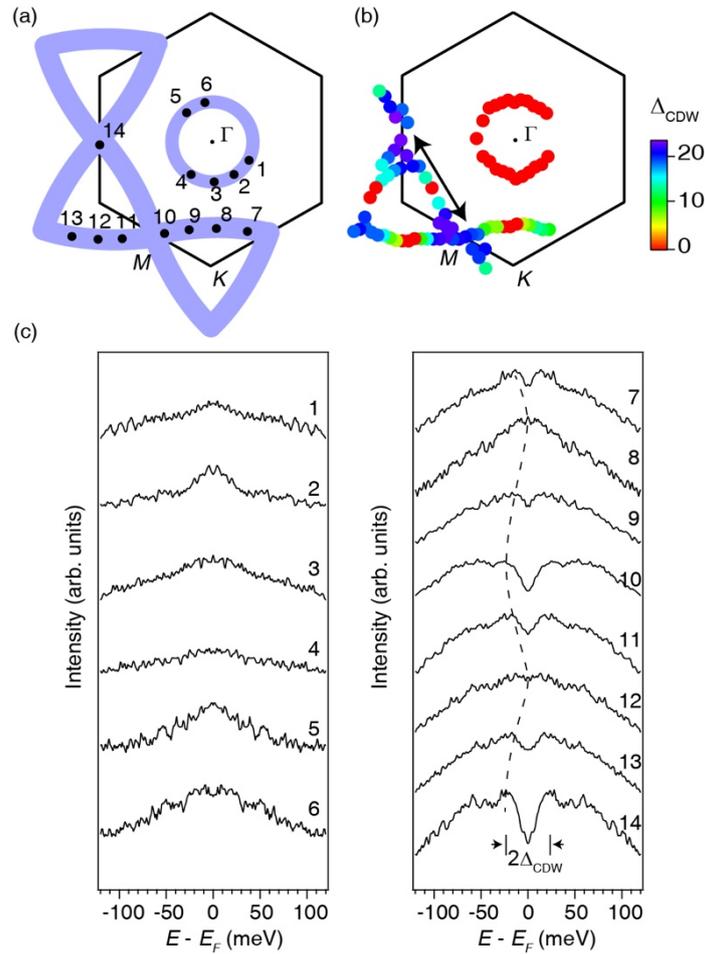

**Figure 4.** (a) Illustration of the Fermi surface topology and the representative momenta. (b) the momentum distribution of the CDW gap on the Fermi surface of $CsV_3Sb_5$. The black arrow illustrates the scattering of electrons between neighboring *M* points. (c) The representative symmetrized EDCs taken at the Fermi crossings that are shown in panel (a). The dashed line guides the eye to the momentum dependence of the CDW gap. All data were taken at 15 K.

Figure 4 shows how the CDW gap distributes in momentum space. We took high-resolution spectra across the Fermi surface of $CsV_3Sb_5$ and calculated $\Delta_{CDW}$ using the energy separation between the two nearest peaks near $E_F$ in the symmetrized EDCs. Some representative symmetrized EDCs are summarized in Fig.4(c). The gap mapping shows that the CDW gap is strongly momentum dependent. The gap maxima locate around the *M* points, while no gap is observed on the Fermi pocket around Γ and the Fermi crossings along the Γ-*K* high symmetry direction.

In a Fermi nesting scenario, CDW is driven by a divergence of electronic susceptibility due to the scattering of electrons between two well nested sections of Fermi surface[23]. For a 2×2 CDW order of CsV$_3$Sb$_5$, considering its Fermi surface topology, the nesting vector in momentum space has two possibilities. One is the Γ-*M* nesting between the Fermi pockets at the Γ and *M* points and the other is the *M*-*M* nesting between the electronic states at neighboring *M* points. Clearly, the momentum distribution of the CDW gap favors the second nesting scenario. According to the band calculations, the band structure of CsV$_3$Sb$_5$ hosts multiple saddle points at the *M* points [3,24]. These saddles points are close to $E_F$, leading to a divergence of density of state. As a result, the scattering of electrons among the saddle points at *M* is strongly enhanced, contributing to a charge instability of Fermi surface which drives the 2×2 CDW order in CsV$_3$Sb$_5$.

We notice that the spectra at $k_{F, M}$ exhibits multi-peak behavior in the CDW state [Fig 3]. Namely, besides the ~20 meV CDW gap, there is another peak at ~60 meV. Previous band calculation and STM tunneling spectrum[14] suggested two Van-Hove singularities (VHS) points around ~36.0 meV (VHS$_1$) and ~76.5 meV (VHS$_2$). Therefore, one simple explanation is that the 20 meV peak comes from the CDW pair-breaking peak associated with the VHS$_1$ band and the 60 meV peak is associated with the VHS$_2$ band which gradually shows up at low temperature owing to weight transfer or further separation between two VHSs. On the other hand, the optical spectroscopy measurements of AV$_3$Sb$_5$ also observed a 60~80 meV gap [25,26], which is comparable with the energy scale of the 60 meV peak. Hence, another possible explanation is that the CDW gap exhibits multi-gap behavior. The 20 meV and 60 meV peaks could be attributed to two separate CDW gaps that open on different bands. However, in this scenario, the 60 meV energy scale seems to be too large, incompatible with a CDW onset temperature of 94 K. We also note that, besides the 2×2 translational symmetry breaking, the CDW state of AV$_3$Sb$_5$ further breaks the time-reversal symmetry or rotational symmetry as proposed previously [12,14-17]. The temperature dependence of the 60 meV peak (peak2) seems to be different from that of the 20 meV peak (peak1), showing an abnormal behavior at ~70 K, which suggests that the high energy peak may correspond to a different symmetry breaking that develops slightly below T* in addition to the CDW state of CsV$_3$Sb$_5$. In a short, further theoretical and experimental efforts are required to pin down the proper scenario for the multi-peak feature.

Finally, we discuss the intriguing coexistence of CDW with superconductivity in AV$_3$Sb$_5$ [27-30]. In most CDW materials, the superconductivity competes strongly with the CDW order in the electronic phase space. Both CDW and SC originate from the scattering of electrons on the Fermi surface. Therefore, they compete with each other to gain the electronic states close to $E_F$. In CsV$_3$Sb$_5$, however, the Fermi surface is only partially gapped in the CDW state. The observation of such strongly momentum-dependent CDW gap provides an important evidence for a microscopic coexistence of the CDW and SC orders in the momentum space. While the scattering of electrons between the neighboring *M* points drives the formation of CDW, the electronic states on the residual Fermi pockets [red points in Fig. 4(b)] may play an important role in the superconducting pairing. Therefore, our results imply an intriguing interplay between the CDW and SC states, and put a strong constraint on constructing an appropriate theoretical model for the CDW and SC mechanisms in this Kagome compound.

The authors acknowledge Prof. F. Zhou and Prof. J. Yuan for valuable discussions. This work is supported by the National Natural Science Foundation of China (Grant Nos. 11888101, 11834016, 12061131005, 51771224, and 61888102), the National Key Research and Development Program of China (Grant Nos. 2018YFA0305602, 2016YFA0301003, 2017YFA0303003, and 2018YFA0305800), the Key Research Program and Strategic Priority Research Program of Frontier Sciences of the Chinese Academy of Sciences (Grant Nos. QYZDY-SSW-SLH001, XDB33010200, and XDB25000000).

and Yang Z 2021, ArXiv:2103.12507